\documentclass[aps,prb,twocolumn,superscriptaddress,longbibliography,showpacs]{revtex4-2}
\usepackage{amsmath,amssymb}
\usepackage{graphicx}
\usepackage{xcolor}

\usepackage{bm}

\definecolor{LinkColor}{rgb}{0.256,0.439,0.588}
\usepackage{hyperref}
\hypersetup{
	pdfauthor={good guys},
	pdftitle={good title},
	colorlinks=true,
	citecolor=LinkColor,
	linkcolor=LinkColor,
	urlcolor=LinkColor
}

\newcommand{\ket}[1]{\left\lvert#1\right\rangle}
\newcommand{\bra}[1]{\left\langle#1\right\rvert}

\begin{document}
\title{Fidelity susceptibility as a diagnostic of the commensurate-incommensurate transition: A revisit of the programmable Rydberg chain}

\author{Xue-Jia Yu}
\altaffiliation{The first two authors contributed equally.}
\affiliation{International Center for Quantum Materials, School of Physics, Peking University, Beijing 100871, China}

\author{Sheng Yang}
\altaffiliation{The first two authors contributed equally.}
\affiliation{Zhejiang Institute of Modern Physics and Department of Physics, Zhejiang University, Hangzhou 310027, China}

\author{Jing-Bo Xu}
\email{xujb@zju.edu.cn}
\affiliation{Zhejiang Institute of Modern Physics and Department of Physics, Zhejiang University, Hangzhou 310027, China}

\author{Limei Xu}
\email{limei.xu@pku.edu.cn}
\affiliation{International Center for Quantum Materials, School of Physics, Peking University, Beijing 100871, China}
\affiliation{Collaborative Innovation Center of Quantum Matter, Beijing, China}
\affiliation{Interdisciplinary Institute of Light-Element Quantum Materials and Research Center for Light-Element Advanced Materials, Peking University, Beijing, China}

\date{\today}

\begin{abstract}
In recent years, programmable Rydberg-atom arrays have been widely used to simulate new quantum phases and phase transitions, generating great interest among theorists and experimentalists. Based on the large-scale density matrix renormalization group method, the ground-state phase diagram of one-dimensional Rydberg chains is investigated with fidelity susceptibility as an efficient diagnostic method. We find that the competition between Rydberg blockade and external detuning produces unconventional phases and phase transitions. As the Rydberg blockade radius increases, the phase transition between disordered and density-wave ordered phases changes from the standard Potts universality class to an unconventional chiral one. As the radius increases further (above Potts point but still close to the tip of the lobe), a very narrow intermediate floating-phase region begins to appear. A concise physical picture is also provided to illustrate the numerical results. Compared with previous studies, this work brings more evidence for commensurate-incommensurate quantum phase transitions in programmable quantum simulators from the perspective of quantum information, showing that fidelity susceptibility can be used to study such phase transitions.
\end{abstract}

\maketitle

\section{Introduction}
\label{sec:introduction}
Mastering exotic phases and correlated phase transitions in quantum many-body systems is one of the core issues in condensed matter physics~\cite{sachdev_2011,sondhi1997}. Unlike phase transitions driven by thermal fluctuations, quantum systems driven only by quantum fluctuations can undergo quantum phase transitions (QPTs) at the absolute zero temperature. By tuning non-thermal parameters such as doping in the parent compound of high-$T_{\rm{c}}$ superconductors~\cite{lee2006rmp}, magnetic field in quantum Hall samples~\cite{stormer1999rmp}, pressure in quantum magnetic systems~\cite{zhou2017rmp,Savary_2016}, and disorder in a conductor near its metal-insulator transition~\cite{evers2008rmp}, the ground state of a quantum system can change fundamentally. Unconventional QPTs have attracted great interest in the past few decades~\cite{xu2012unconventional,Guo2022pra,yu2022emergent,yu2021conformal} and have raised new questions both theoretically and experimentally, especially the absolute necessity to consider quantum effects~\cite{sachdev_2011}.

The concept of universality classes plays a crucial role in the study of continuous phase transitions~\cite{cardy_1996}. Quantum many-body systems with nearest-neighbor interactions, such as transverse-field Ising model, Heisenberg model, and Hubbard model, are crucial for understanding QPTs~\cite{sachdev_2011}. By constructing simplified lattice models, the same low-energy physics can be studied in terms of the concept of universality, even if we use different microscopic lattice Hamiltonians. Furthermore, many quantum critical points are marked by the emergence of scale invariance or conformal symmetry, and the corresponding universality class can be described by conformal field theories (CFTs) with a dynamical critical exponent $z=1$~\cite{cardy_1996,francesco2012conformal,ginsparg1988applied}.

Recently, neutral Rydberg atoms trapped in optical tweezers with programmable van der Waals type interactions~\cite{keesling2019quantum,bernien2017probing,ebadi2021quantum,browaeys2020many} provide a promising tunable platform for observing various quantum phenomena, such as gapped $\mathbb{Z}_{2}$ quantum spin liquid~\cite{ruben2021prx,semeghini2021probing,giudici2022dynamical,yan2022triangular,slagle2022quantum,samajdar2022emergent,giudice2022trimer,ruben2205,Lee_arxiv_2022}, QPTs between different density-wave-ordered and disordered phases~\cite{an2022quantum,o2022entanglement,marcin2022prb,samajdar2021quantum,samajdar2020prl,slagle2021prb,chandran2020prb,giudici2019prb}, quantum Kibble-Zurek (KZ) mechanism~\cite{keesling2019quantum,chepiga2021kibble,huang2019prb}, and unexpected quantum many-body scars~\cite{turner2018weak,serbyn2021quantum}. However, for long-range quantum many-body systems, it remains challenging to fully understand their critical behaviors, either through theoretical analyses or numerical simulations. As a prototype example, QPTs in Rydberg chains are even more subtle. The quantum Kibble-Zurek experiments~\cite{keesling2019quantum,chepiga2021kibble,huang2019prb} dynamically probed QPTs between an incommensurate (IC) disordered phase and a commensurate (C) ordered phase with period $p=3,4,\dots$, sparking interest in the C-IC transition first proposed in the context of adsorbed monolayers in $1980$s and $1990$s ~\cite{ostlund1981prb,huse1981prb,huse1982prl,huse1984prb}. The C-IC transition with $p\ge5$ emerges through an intermediate gapless floating phase (the Luttinger liquid) with a central charge $c=1$, characterized by incommensurate correlations and wave vector $q$~\cite{maceira2022conformal,NYCKEES2021115365,nyckees2022prr,chepiga2019prl,rader2019floating}. On the one hand, the disordered-to-floating transition belongs to the Kosterlitz-Thouless (KT) universality class with an exponentially diverged correlation length. On the other hand, the transition from the floating phase to the commensurate ordered one is a Pokrovsky-Talapov (PT) transition.


In this work, we use the concept of fidelity susceptibility borrowed from quantum information theory~\cite{gu2010fidelity,albuquerque2010prb} to provide a new perspective on the C-IC transition. As a purely geometric measure of quantum states, fidelity susceptibility is believed to be effective in characterizing sudden changes in the ground-state structure associated with QPTs, and over the past few years, this concept has been established as one of the powerful diagnostic methods for QPTs without prior knowledge of order parameters or associated symmetry-breaking patterns. To date, fidelity susceptibility has been applied to detect various quantum critical points, such as conventional symmetry-breaking quantum critical points~\cite{zhu2018pra,sun2017pra}, topological phase transitions~\cite{sun2015prb}, Anderson transitions~\cite{wei2019pra,lv2022pra}, deconfined quantum criticality~\cite{sun2019prb}, and even non-Hermitian critical points~\cite{sun2022biorthogonal,chin2021prr,tu2022general}. In this work, we will show that this concept could also be an attractive tool for studying challenging C-IC problems. Specifically, the fidelity susceptibility of QPTs can be experimentally detected by neutron scattering or the angle-resolved photoemission spectroscopy techniques~\cite{gu2010fidelity,gu2014epl}. 

The rest of this paper is organized as follows: Sec.~\ref{sec:model} contains a brief introduction to the lattice model of the Rydberg-atom chain and the numerical method adopted. A concise review of previous results about the C-IC transition, the concept of fidelity susceptibility, and relevant finite-size scaling behaviors are also given therein. Then a standard finite-size scaling analysis is applied to study the transition from the period-2 ordered phase to the disordered phase which belongs to the Ising universality in Sec.~\ref{sec:ising}. The intermediate floating phase or exotic chiral transition between the disordered phase and the period-3 ordered phase is then explored with the same approach, and a brief explanation of the numerical results is provided in Sec.~\ref{sec:chiral}. Finally, we give a conclusion in Sec.~\ref{sec:summary}. Additional data of our numerical calculations are provided in Appendixes.

\section{Model and method}
\label{sec:model}

\subsection{Rydberg Hamiltonian}
In this work, we study the Hamiltonian of interacting Rydberg atoms arranged in a one-dimensional chain of length $L$ with open boundary conditions~\cite{samajdar2020prl}, 
\begin{equation}
\begin{split}
\label{eq:hamiltonian}
H_{\rm{Ryd}} = & \sum_{i=1}^{L}\left[\frac{\Omega}{2}\left(\ket{r}_{i}\!\bra{g}+\ket{g}_{i}\!\bra{r}\right) - \delta \ket{r}_{i}\!\bra{r}\right] \\
& + \sum_{i<j}V(|i-j|) \ket{r}_{i}\!\bra{r}\otimes\ket{r}_{j}\!\bra{r}\,.
\end{split}
\end{equation}
Here, $i$ represents the discrete sites of the Rydberg lattice (with lattice constant $a$). $\ket{g}_{i}$ and $\ket{r}_{i}$ denote the internal atomic ground state and an excited Rydberg state of the $i$th atom, respectively. The Rabi frequency $\Omega$ and detuning $\delta$ characterize a coherent laser driving field. $V(|i-j|) = C_{6}/|i-j|^{6}$ is the strength of the van der Waals interaction of atoms excited to the Rydberg state $\ket{r}$. The long-range interactions here can also be equivalently parametrized by the Rydberg blockade radius $R_{\rm{b}}$, defined by $V(R_{\rm{b}}/a)\equiv\Omega$~\cite{samajdar2020prl}. Where interactions are so strong that the Rydberg excitation of one atom suppresses the excitation of other nearby atoms. This effect is called the Rydberg blockade mechanism. Notably, the model Hamiltonian $H_{\rm{Ryd}}$ can be mapped to a hard-core boson model by identifying $\ket{g}$ and $\ket{r}$ as the empty state and occupied state, respectively~\cite{fendley2004prb,subir2002prb}. For simplicity, $\Omega$ and $a$ are set to energy and length units in our actual numerical simulations.

\begin{figure}[tb]
\includegraphics[width=\linewidth]{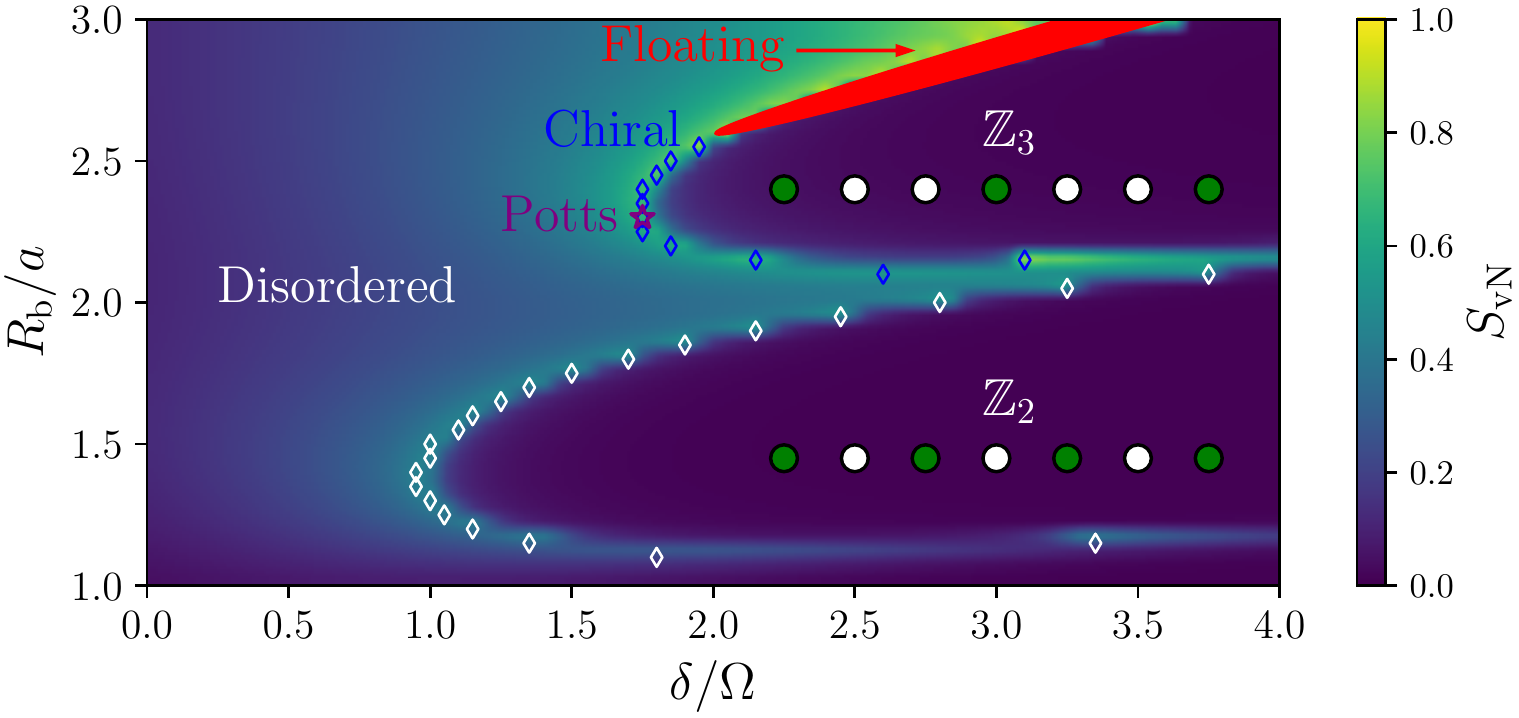}
\caption{
The ground-state phase diagram of Rydberg chains is mapped out with respect to the Rydberg blockade radius $R_{\rm{b}}/a$ and detuning $\delta/\Omega$ by exploiting the half-chain entanglement entropy~\cite{nielsen_chuang_2010,liu2016prl,samajdar2020prl}, $S_{\rm{vN}}\equiv -{\rm{Tr}}\,[\rho_{l(r)}\ln\rho_{l(r)}]$, with an open-chain length $L=121$; $\rho_{l(r)}$ is the reduced density matrix of the left $60$ (right $61$) sites of the chain. $\mathbb{Z}_{n}$ ($n=2,3$) represents the $\mathbb{Z}_{n}$ symmetry-breaking ordered phase with green (white) circles are atoms in the Rydberg state $\lvert{r}\rangle$ (the electronic ground state $\lvert{g}\rangle$). The phase boundaries are roughly estimated by the local maximums of $S_{\rm{vN}}$~\cite{liu2016prl}. The purple point refers to the phase transition belonging to the three-state Potts universality class. The blue points represent the non-conformal chiral phase transition. The red parameter region \textit{schematically} indicates the potential intermediate floating phase according to recent related works~\cite{rader2019floating,chepiga2019prl,maceira2022conformal} (also see the discussion in Sec.~\ref{sec:chiral}). It is noted that there is also a floating region below the $\mathbb{Z}_{3}$ lobe which is beyond the parameter range shown here (far away from the lobe tip)~\cite{rader2019floating}. Simulation parameters: MPS bond dimension is $500$, the relative energy error is smaller than $10^{-10}$, the parameter resolution is ${\rm{d}}(R_{\rm{b}}/a)={\rm{d}}(\delta/\Omega)=0.05$\,.
}
\label{fig:phase_diagram}
\end{figure}

The ground state of the Rydberg Hamiltonian, $\ket{\phi}$, depends sensitively on the detuning $\delta/\Omega$ and the blockade radius $R_{\rm{b}}$, which govern the density of Rydberg excitation $n\equiv\bra{\phi}(\sum_{i}\ket{r}_{i}\!\bra{r}/L)\ket{\phi}$. At large negative $\delta/\Omega$, it is favorable for most atoms to be in the electronic ground state $\ket{g}$, which corresponds to the disordered phase. Whereas for large positive values of $\delta/\Omega$, the Rydberg excitation density $n$ increases, and due to the Rydberg blockade mechanism, complex density-wave ordered phases with different spatial symmetries will be established depending on $R_{\rm{b}}/a$, called ``Rydberg crystal" phases (see Fig.~\ref{fig:phase_diagram}).

\subsection{Numerical method}
Since there is no exact analytical solution for the general Rydberg Hamiltonian $H_{\rm{Ryd}}$, we employ a large-scale finite-size density matrix renormalization group (DMRG) method~\cite{white1992prl,SCHOLLWOCK201196} in the representation of matrix product states (MPS)~\cite{perez2006matrix}, which is one of the most powerful numerical methods today for solving the ground states of 1D strongly correlated many-body systems. More specifically, to reduce the computational requirement due to the long-range interaction term $V(r)$, we employ a truncation strategy similar to that used in Ref.~\cite{samajdar2018pra} (see the U-V model~\cite{fendley2004prb} therein). In real numerical simulations, we preserve interactions at most fourth-nearest neighbors by forcing $V(r>4)=0$. Compared to the U-V model, the Hamiltonian with the interaction truncation adopted here is more relevant to actual experiments and can more faithfully simulate the effective physics of the ordered-to-disordered phase transition in the Rydberg chain.

We note that the period-$p$ ordered-to-disordered phase transitions of the model will be studied together with fixed $R_{\rm{b}}/a$ lines, so the Hamiltonian therefore can be viewed as a function of detuning. Fidelity susceptibility can then be calculated by~\cite{chen2008pra,you2007pre} 
\begin{equation}
\label{eq:fidelity}
\chi_{F}(\delta)=\lim_{{\rm{d}}\delta\rightarrow0}\frac{2[1-\lvert\langle{\phi(\delta)}\vert{\phi(\delta+{\rm{d}}\delta})\rangle\rvert]}{({\rm{d}}\delta)^{2}}\,,
\end{equation}
where $\ket{\phi(\delta)}$ is the ground state of $H_{\rm{Ryd}}(\delta)$. The quantum fidelity~\cite{nielsen_chuang_2010}, $F(\delta,{\rm{d}}\delta)=\lvert\langle{\phi(\delta)}\vert{\phi(\delta+{\rm{d}}\delta})\rangle\rvert$, is defined as the overlap amplitude of two nearby quantum states in the parameter space. In general, the fidelity is expected to drop sharply at quantum critical points, reflecting a sudden change in the ground-state structure. Therefore, the susceptibility $\chi_{F}(\delta)$, which is the dominant quadratic term in the series expansion with respect to ${\rm{d}}\delta$, can faithfully diagnose general QPTs. Specifically,  fidelity susceptibility has been successfully used to study various types of QPTs, including conventional~\cite{zhu2018pra,sun2017pra} and unconventional ones~\cite{sun2015prb,sun2019prb}. We emphasize that the application of this method does not rely on any prior knowledge of the underlying order parameters. 

For continuous QPTs, it has been established in the literature that the fidelity susceptibility obeys certain finite-size scaling laws near quantum critical points. This paves the way for extracting relevant key indices about critical exponents. In particular, the fidelity susceptibility exhibits a universal dependence on the control parameter described by the functional form~\cite{gu2010fidelity,albuquerque2010prb,gu2008prb} 
\begin{equation}
\label{eq:fs_collapse}
\chi_{F}(\delta, L) = L^{2/\nu}\mathcal{F}_{F}\big[L^{1/\nu}(\delta-\delta_{\rm{c}})\big]\,,
\end{equation}
where $\nu$ is the correlation length exponent and $\mathcal{F}_{F}$ is an unknown scaling function. Furthermore, for finite chains, the fidelity susceptibility at the pseudocritical point $\delta_{\rm{m}}$ also shows a power-law behavior with respect to the chain length $L$, 
\begin{equation}
\label{eq:fs_scaling}
\chi_{F}(\delta_{\rm{m}}, L) \propto L^{2/\nu}\,.
\end{equation}
We note that some sub-leading terms usually on the right-hand side of this relation could be ignored if the system size is large enough. Based on the above two scaling forms, it is easy to obtain the value of $\nu$ for the QPT we are interested in here and to determine whether it is continuous or not. 

To further reveal the nature of the conformal or non-conformal phase transitions, we also calculate the energy gap $\Delta$, which is defined as the energy difference between the first excited state and the ground state of $H_{\rm{Ryd}}$, to obtain the dynamical critical exponent $z$. For continuous phase transitions, the energy gap is expected to disappear with $\Delta\sim\lvert{\delta-\delta_{\rm{c}}}\rvert^{z\nu}$ as $\delta$ approaches $\delta_{\rm{c}}$~\cite{sachdev_2011}. Combined with the divergence of the correlation length following the form, $\xi\sim\lvert{\delta-\delta_{\rm{c}}}\rvert^{-\nu}$, we obtain the scaling relation, $\Delta\sim\xi^{-z}$. Since the correlation length at the critical point of a finite system can be characterized by the lattice length $L$, the finite-size scaling form, $\Delta(\delta_{\rm{m}}, L)\propto L^{-z}$, can be finally derived. In addition, the energy gap also exhibits a similar functional form to the fidelity susceptibility~\cite{samajdar2018pra} 
\begin{equation}
\label{eq:ge_collapse}
\Delta(\delta, L) = L^{-z}\mathcal{F}_{\Delta}\big[L^{1/\nu}(\delta-\delta_{\rm{c}})\big]\,,
\end{equation}
where $\mathcal{F}_{\Delta}$ is another scaling function associated with $\Delta$. We note that the dynamical critical exponent  $z$ equals to $1$ for conformal universality classes. To unveil the chiral nature of the QPTs, however, we still have to resort to the diagnosis proposed originally by Huse and Fisher~\cite{huse1982prl}, which is described and discussed in Appendix~\ref{sec:chiralcheck}.

\subsection{Brief review of previous results}
The transition out of a period-$p$ phase is an example of the C-IC transition, a problem with a long history that dates back to the investigation of adsorbed monolayers on surfaces~\cite{bonn1994prb,schreiner1994prb}. Naively, it is expected that such C-IC transitions should belong to the standard Potts universality class. However, as first suggested by Huse and Fisher~\cite{huse1981prb}, the system will introduce a chiral perturbation if different phases have inequivalent domain walls. If this perturbation is relevant, the standard Potts universality class can only exist at an isolated point where the perturbation vanishes. Away from the Potts point, there is a question under debate during the past decades: what is the nature of the C-IC transition? There are three possibilities: (i) there is still a unique transition, but it belongs to a non-conformal chiral universality class~\cite{huse1982prl,huse1984prb}; (ii) there is a critical Luttinger-liquid intermediate phase called the floating phase~\cite{haldane1983prb,schulz1983prb}; (iii) the transition is first order. 

These issues were raised again by Fendley et al.~\cite{fendley2004prb,subir2002prb}. It is also related to recent Rydberg-atom experiments in the context of one-dimensional quantum models of constraint bosons. More recently, some groups~\cite{samajdar2018pra,whitsitt2018prb} used numerical and field-theory methods to show that there is a direct continuous chiral phase transition between the disordered phase and the period-3 ordered phase, without the intermediate floating phase. However, some other groups~\cite{chepiga2019prl,rader2019floating} performed large-scale state-of-art numerical calculations to provide strong evidence that the intermediate floating phase can occur sufficiently far from the Potts point, confirmed by the extrapolation of correlation lengths and incommensurate wave vectors. Furthermore, a theoretical argument was also presented in Ref.~\cite{mila2021prr} for the existence of a Lifshitz point, which draws a clear line between the chiral transition and the floating phase for $p=3,4$ C-IC transitions. 

Notably, equipped with a newly developed tensor network method~\cite{chepiga2019prl}, the system size considered in previous calculations reaches up to 9000 sites, and the scaling properties of the wave vector and the correlation length around the critical points can be precisely obtained. It will be interesting to see if there is any other physical quantity that can efficiently identify the possible unconventional phases or phase transitions with relatively smaller system sizes. Therefore, it is worth revisiting this period-3 ordered-to-disordered phase transition from a different perspective, which will address these issues in the next few sections.

\section{Warm-up: Ising universality class}
\label{sec:ising}
In the following, based on the scaling laws listed above, we first consider the period-2 ordered-to-disordered transition, which belongs to the (1+1)D Ising universality class~\cite{sachdev_2011} and has been well studied. This allows us to test the feasibility of the fidelity susceptibility and energy gap approach to determine the critical exponents $\nu$ and $z$.

To avoid defects induced by edge excitation and thus stabilize the ordered phase in the bulk, we consider system sizes $L=2n+1$, ranging from $L=49$ to $129$ sites for the period-2 ordered-to-disordered transition in our DMRG simulations with open boundary conditions. The MPS bond dimension is set to 300; when calculating the ground and the first excited states, a good convergence to the true energy eigenstates is guaranteed by requiring the relative energy error to be smaller than $10^{-10}$ and $10^{-8}$, respectively. For the fidelity susceptibility, we set a stricter convergence criterion with a relative energy error lower than $10^{-12}$, and the detuning step is chosen as ${\rm{d}}\delta=10^{-3}$ [see Eq.~\eqref{eq:fidelity}].

\begin{figure}[tbp]
\includegraphics[width=1.\linewidth]{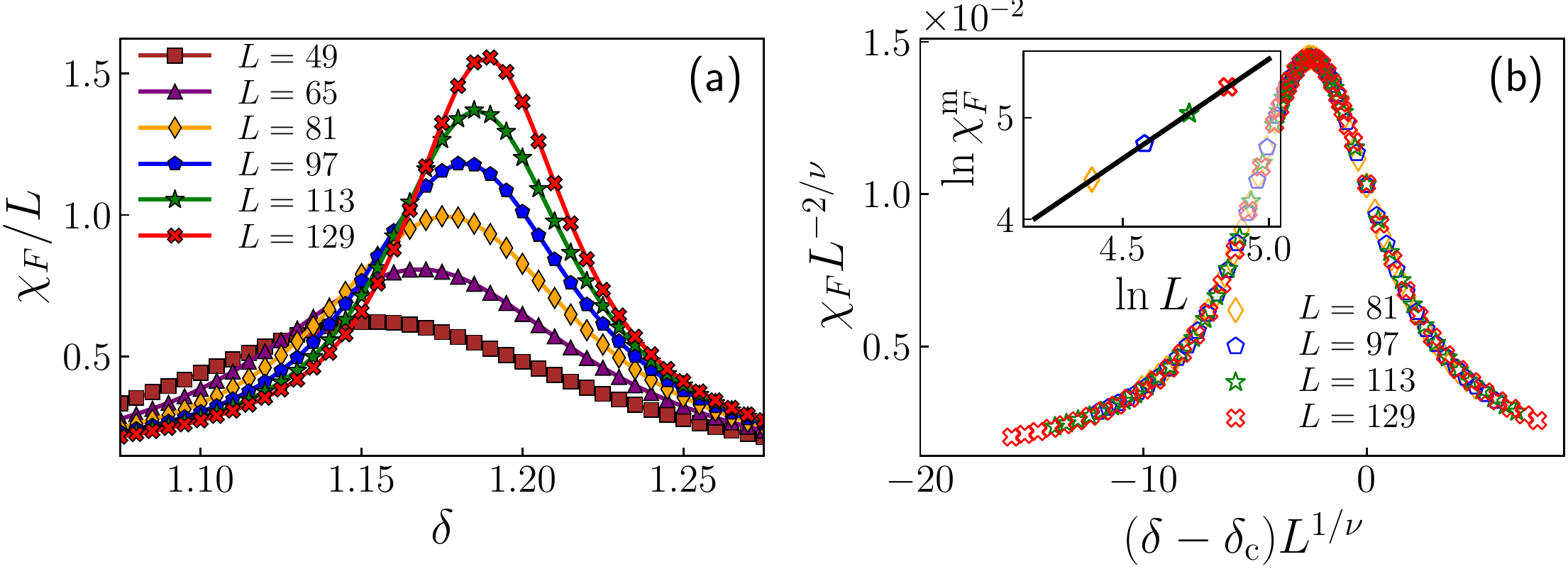}
\caption{
The finite-size scaling analysis of the fidelity susceptibility for the period-2 ordered-to-disordered transition with a fixed blockade radius $R_{\rm{b}}/a=1.6$\,. (a) The fidelity susceptibility per site shows a sharp peak near the transition point. (b) Data collapse of the rescaled fidelity susceptibility and detuning with $\nu=1.019$ and $\delta_{\rm{c}}=1.210$ for various system sizes. The inset shows the log-log plot of the fidelity susceptibility against the system size at the pseudocritical point, and the correlation length critical exponent $\nu=1.019$ can be inferred from the slope of the fitted straight line.
}
\label{fig:fs1.6}
\end{figure}

Fig.~\ref{fig:fs1.6}(a) shows the fidelity susceptibility per site $\chi_{F}/L$ as a function of detuning $\delta$, with a fixed blockade radius $R_{\rm{b}}/a=1.6$\,. It is clear that $\chi_{F}/L$ exhibits a sharp peak near the critical point and the divergence behavior can be described by conventional scaling laws, as expected from Eqs.~\eqref{eq:fs_collapse} and~\eqref{eq:fs_scaling}. To extract the value of $\nu$, we calculate more data points around the peak for each system size to obtain fidelity susceptibility $\chi_{F}^{\rm{m}}$ at the finite-size pseudocritical point $\delta_{\rm{m}}$. The inset of Fig.~\ref{fig:fs1.6}(b) exhibits a linear fit to the DMRG data of the largest four system sizes using the least square method, consistent with a linear correlation of the logarithm of $\chi_{F}^{\rm{m}}$  to $\ln{L}$ based on Eq.~\eqref{eq:fs_scaling}. An accurate estimation of the critical exponent $\nu=1.019$ can then be easily acquired from the slope of the fitted straight line. On the other hand, the presence of the scaling function $\mathcal{F}_{F}$ [see Eq.~\eqref{eq:fs_collapse}] usually provides independent verification of the correctness of the critical exponents. Using the value of $\nu$ obtained from the log-log plot and fine tuning of the critical point $\delta_{\rm{c}}$ (considered here as a tunable variable), we use the rescaled variables $\chi_{F}L^{-2/\nu}$ and $(\delta-\delta_{\rm{c}})L^{1/\nu}$ for various system sizes $L$. By implementing a good data collapse, as shown in Fig.~\ref{fig:fs1.6}(b), the critical point can eventually be located at $\delta_{\rm{c}}=1.210(2)$; the associated uncertainty is determined from the visible imperfection observed in the process.

\begin{figure}[tbp]
\includegraphics[width=1.\linewidth]{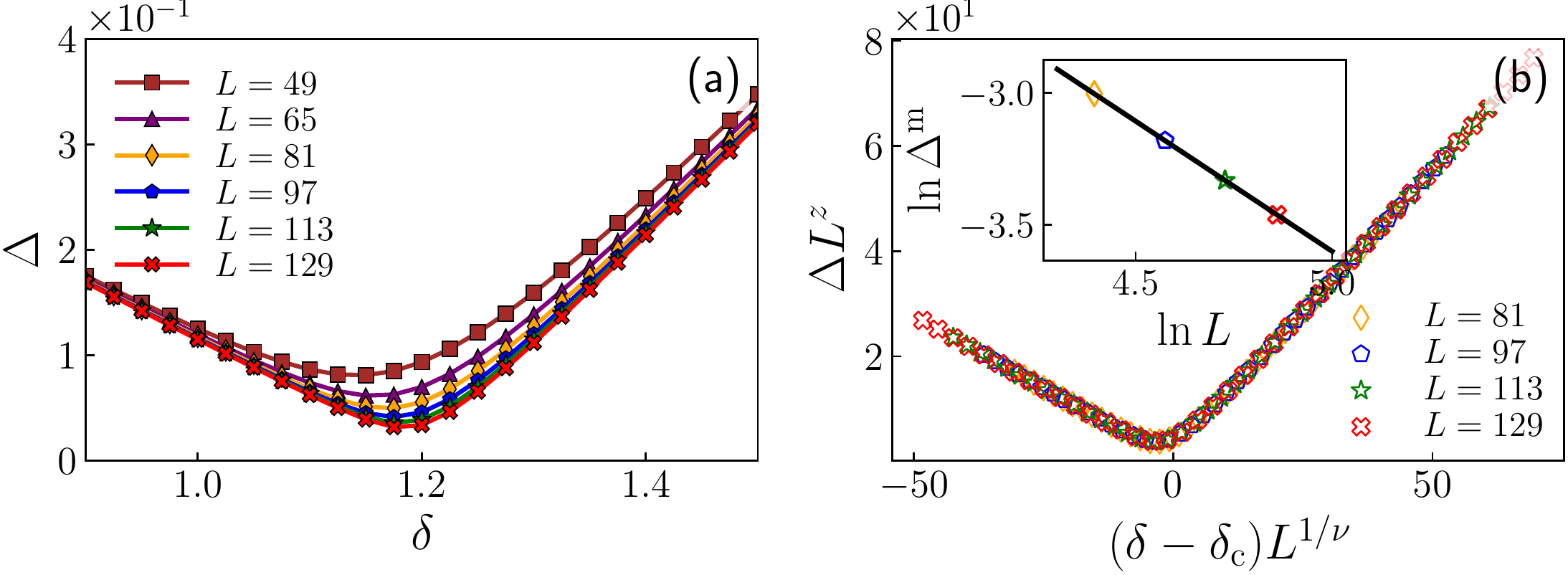}
\caption{
The finite-size scaling analysis of the energy gap $\Delta$ for the period-2 ordered-to-disordered transition with a fixed Rydberg blockade radius $R_{\rm{b}}/a=1.6$\,. (a) The energy gap develops a deep valley near the transition point. (b) Data collapse of the rescaled energy gap and detuning with $z=0.9847$, $\nu=1.019$, and $\delta_{\rm{c}}=1.210$ for the largest four system sizes. The inset displays the log-log plot of the energy gap $\Delta^{\rm{m}}$ versus the system size $L$ at the pseudocritical point $\delta_{\rm{m}}$, and the fitted straight line has a slope whose absolute value equals to the dynamical critical exponent $z$.
}
\label{fig:eg1.6}
\end{figure}

We now obtain $\delta_{\rm{c}}$ and $\nu$ via a standard finite-size scaling analysis of fidelity susceptibility. However, a single critical exponent, such as $\nu$, is not sufficient to determine the universality class to which a QPT belongs. Therefore, we also acquire the dynamical critical exponent $z$ from a similar scaling analysis of the energy gap $\Delta$. The two-fold degeneracy of the ground state in the period-2 ordered phase is split here due to the addition of an extra site on the right edge, and the energy gap should show a deep valley near the transition point [see Fig.~\ref{fig:eg1.6}(a)] rather than vanishing to zero by entering into the ordered phase. Following a similar logic followed in the fidelity susceptibility analysis, we first estimate the critical exponent $z$ from the log-log plot of the gap $\Delta^{\rm{m}}$ at the pseudocritical point $\delta_{\rm{m}}$ versus system size $L$. It is evident from the inset of Fig.~\ref{fig:eg1.6}(b) that the logarithm of $\Delta^{\rm{m}}$ shows a perfect linear dependence on $\ln{L}$, the absolute value of whose slope should equal to $z$. By exploiting the least square method, we end up with an estimation of $z=0.9847$, which is very close to 1. Finally, we also use the scaling function $\mathcal{F}_{\Delta}$ to confirm the accuracy of the $z$ estimation. By directly inserting $z=0.9847$ obtained from the log-log plot as well as $\nu=1.019$ and $\delta_{\rm{c}}=1.210$ inferred from the fidelity susceptibility into Eq.~\eqref{eq:ge_collapse}, a good collapse of curves [Fig.~\ref{fig:eg1.6}(a)] can be established [Fig.~\ref{fig:eg1.6}(b)] without any free parameters. Note that the values of $\nu$ and $\delta_{\rm{c}}$ used in this collapse are obtained independently from the fidelity method, so such a full collapse also shows that the fidelity susceptibility and energy gap methods agree with each other.

Together with the value obtained so far for the critical exponents $\nu=1.019$ and $z=0.9847$, we can now see within reasonable numerical precision that this result agrees with the well-known results in (1+1)D Ising universality class for $\nu=1$ and $z=1$. We attribute the small difference to a potential finite-size effect; considering larger system sizes can reduce this bias in a controllable manner. Thus, our results strongly demonstrate that the transition between the disordered phase and the period-2 ordered phase realized in Rydberg chains belongs to the (1+1)D Ising universality class. In Appendix~\ref{sec:isingmore}, we also provide a finite-size scaling analysis of the same phase transition but with another Rydberg blockade radius $R_{\rm{b}}/a=1.4$, which leads to the same conclusion.

\section{Non-conformal critical point: chiral universality class}
\label{sec:chiral}
As can see from Sec.~\ref{sec:ising}, the combination of fidelity susceptibility and energy gap methods provides a very powerful and self-consistent way to determine critical exponents and the universality class of QPTs. We, therefore, use this approach and follow the same logic used above in analyzing the period-2 ordered-to-disordered transition, to investigate the more complicated period-3 ordered-to-disordered transition hosted by the programmable Rydberg chain. 

As a manifestation of the C-IC problem in 1D quantum systems, the period-3 ordered-to-disordered transition realized in the Rydberg chain is an ideal place to explore the nature of C-IC transitions, though challenging. In recent years, both theoretical and experimental works focusing on interacting Rydberg chains have made some exciting progress in this direction~\cite{chepiga2019prl,mila2021prr,maceira2022conformal,samajdar2018pra,whitsitt2018prb}. On the one hand, by adiabatically driving a Rydberg chain consisting of 51 neutral atoms through a possible period-3 ordered-to-disordered transition, researchers experimentally measured the corresponding KZ exponent $\mu$ in relation to other critical exponents via the formula $\mu=\nu/(1+z\nu)$~\cite{keesling2019quantum}. On the other hand, numerical studies~\cite{samajdar2018pra} utilizing exact diagonalization and finite-size scaling methods have also extracted the dynamical critical exponent $z$ in some parameter regimes. It can be observed that the critical exponent $z$ varies continuously with the coupling strength (or equivalently, the blockade radius), recovering an exponent value similar to that of the $\mathbb{Z}_{3}$ chiral clock model. This behavior suggests that in  addition to the three-state Potts criticality with $z=1$, a direct continuous chiral phase transition of $z\neq1$ also occurs. Notably, independent evidence of non-conformal chiral transitions was later provided for longer Rydberg chains~\cite{maceira2022conformal} and infinite chains~\cite{rader2019floating}, as well as a potential intermediate floating phase which is only theoretically predicted between the incommensurate disordered and commensurate ordered phases. All these studies expand our understanding of the C-IC transition and motivate us to further work in this research direction. 

This prompts us to revisit the period-3 ordered-to-disordered transition in the Rydberg chain from a different perspective, namely, the fidelity susceptibility, as an independent exploration of this exotic transition. For the same reasons explained in the previous section, we consider system size $L=3n+1$, ranging from $L=49$ to $127$, to stabilize the period-3 ordering in the system bulk. We also note the truncation strategy adopted in our simulations concerning the long-range interaction (see Sec.~\ref{sec:model}). The DMRG-related parameters set here are the same as in the previous section. 

\begin{figure}[tbp]
\includegraphics[width=1.\linewidth]{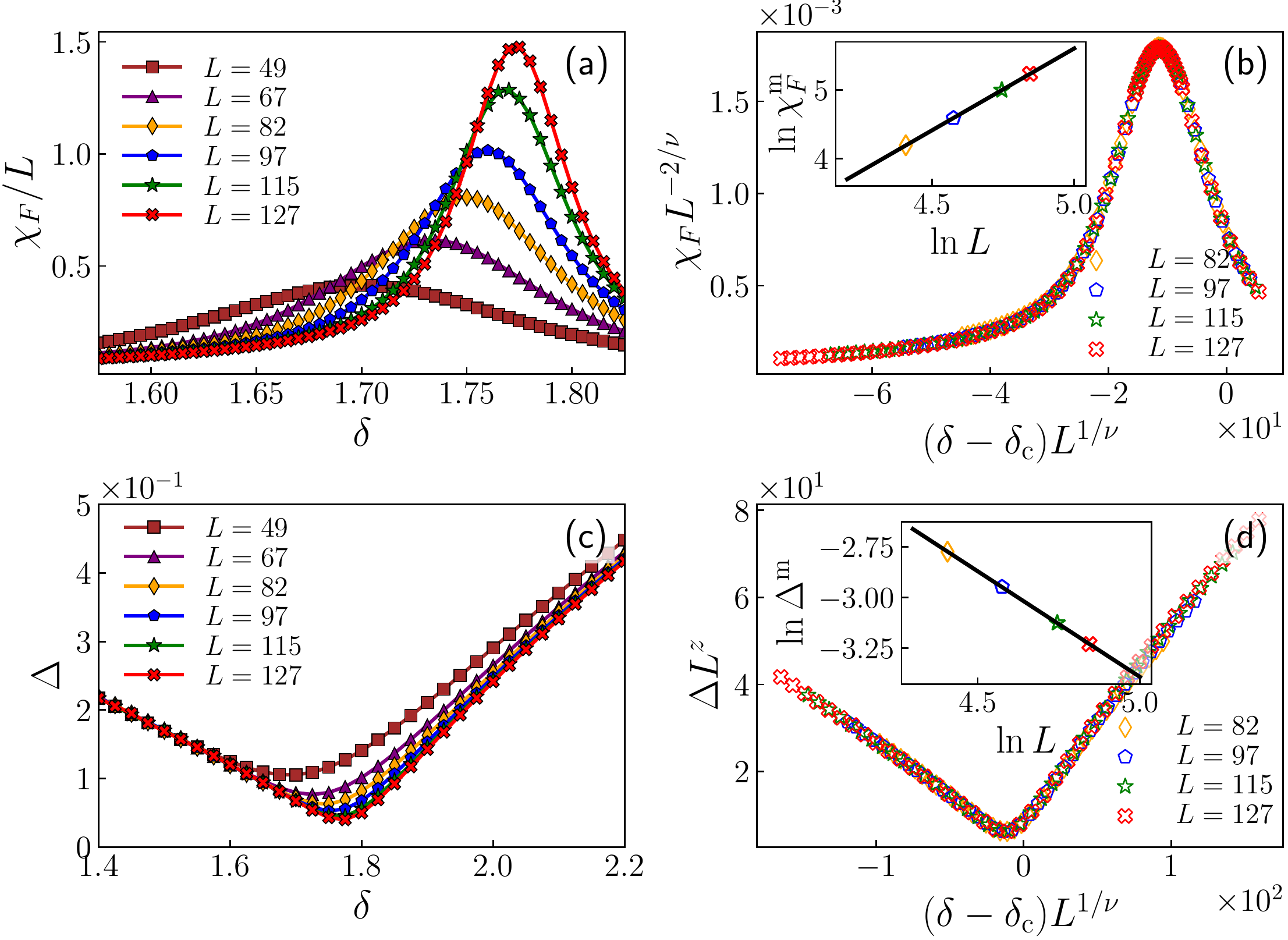}
\caption{
The finite-size scaling analysis of the fidelity susceptibility and energy gap for the period-$3$ ordered-to-disordered transition with $R_{\rm{b}}/a=2.3$\,. The fidelity susceptibility per site (a) and energy gap (c) develop a sharp peak and a deep valley, respectively, near the transition point. A standard finite-size scaling method has been used to estimate the critical point $\delta_{\rm{c}}=1.808$\,, as well as the critical exponents $\nu=0.838$ and $z=1.044$\,.
}
\label{fig:rb2.3}
\end{figure}

We first investigate the case of $R_{\rm{b}}/a=2.3$ near the three-state Potts universality class by possessing critical exponents that is very close to the theoretical result for the universality class. Similar to the performance observed in the period-2 ordered-to-disordered transition, it is evident from Fig.~\ref{fig:rb2.3} that the fidelity susceptibility per site and the energy gap exhibit sharp peaks and deep valleys, respectively, around a certain detuning value signaling the occurrence of possible QPTs.

After a standard finite-size scaling analysis following the same procedure described in Sec.~\ref{sec:ising}, the associated critical exponents can be numerically estimated as $\nu=0.838$ and $z=1.044$, which is consistent with the results for the three-state Potts universality class within numerical precision. In this sense, we can say that the case of $R_{\rm{b}}/a=2.3$ is very close to the exact three-state Potts critical point. It is worth noting that the precise Potts point can be located numerically with high precision by following the commensurate line $q=2\pi/3$ in the disordered phase until the period-3 ordered phase is reached. More details can be found in references~\cite{maceira2022conformal,NYCKEES2021115365,nyckees2022prr,chepiga2019prl}. Furthermore, we also examined the effect of the interaction cutoff adopted in our simulation on this $R_{\rm{b}}/a=2.3$ case. By further considering the fifth-nearest neighbor interaction $V(r=5)$, we find the critical point exhibits a small shift of order $10^{-2}$ to the disordered region. This implies that the truncation $V(r>4)=0$ is a good approximation for the actual long-range coupling; the critical point estimated here should be comparable to the one measured in real experiments.

\begin{figure}[tbp]
\includegraphics[width=1.\linewidth]{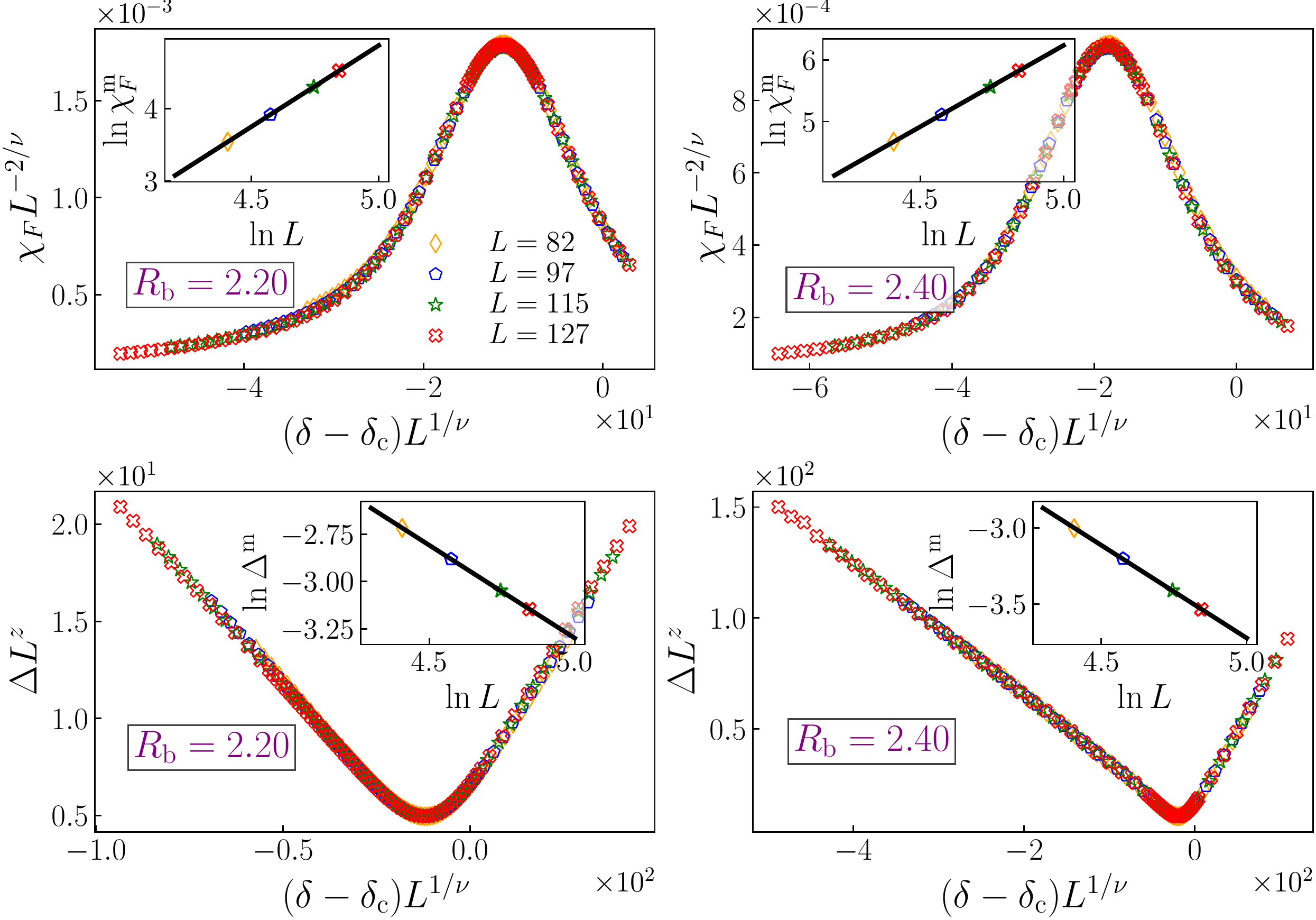}
\caption{
Data collapses of the fidelity susceptibility and energy gap for blockade radii $R_{\rm{b}}/a=2.20$ and $2.40$\,,  within the period-3 ordered-to-disordered transition regime based on the values of $\nu$ and $z$ summarized in Table~\ref{tab:exponents}. Data for different system sizes are denoted by different symbols: orange diamond for $L=82$, blue pentagon for $L=97$, green star for $L=115$, and red cross for $L=127$, as shown in the first plot (same for other plots). The value of the associated critical exponents is estimated, following the same logic obeyed in Sec.~\ref{sec:ising}, by the log-log plot of $\chi_{F}$ and $\Delta$ versus the chain length $L$ at the pseudocritical point (see the insets of the plots).
}
\label{fig:combine_one}
\end{figure}

To reveal the characteristics of direct chiral transitions, we next consider other blockade radius within the period-3 ordered-to-disordered transition regime. As shown in Fig.~\ref{fig:combine_one} and Appendix~\ref{sec:z3more}, both the fidelity susceptibility and energy gap obey the conventional finite-size scaling law over a relatively wide blockade range. In Appendix~\ref{sec:finiteeffect}, we have also included two smaller system sizes $L=49$ and $67$ in the data-collapse plot for the case of $R_{\rm{b}}/a=2.4$ to investigate the effect of the finite-system size on the scaling analysis. It is observed that the system sizes $L$ from $82$ to $127$ are sufficient to acquire reliable critical-exponent estimations.

\begin{figure}[tbp]
\includegraphics[width=\linewidth]{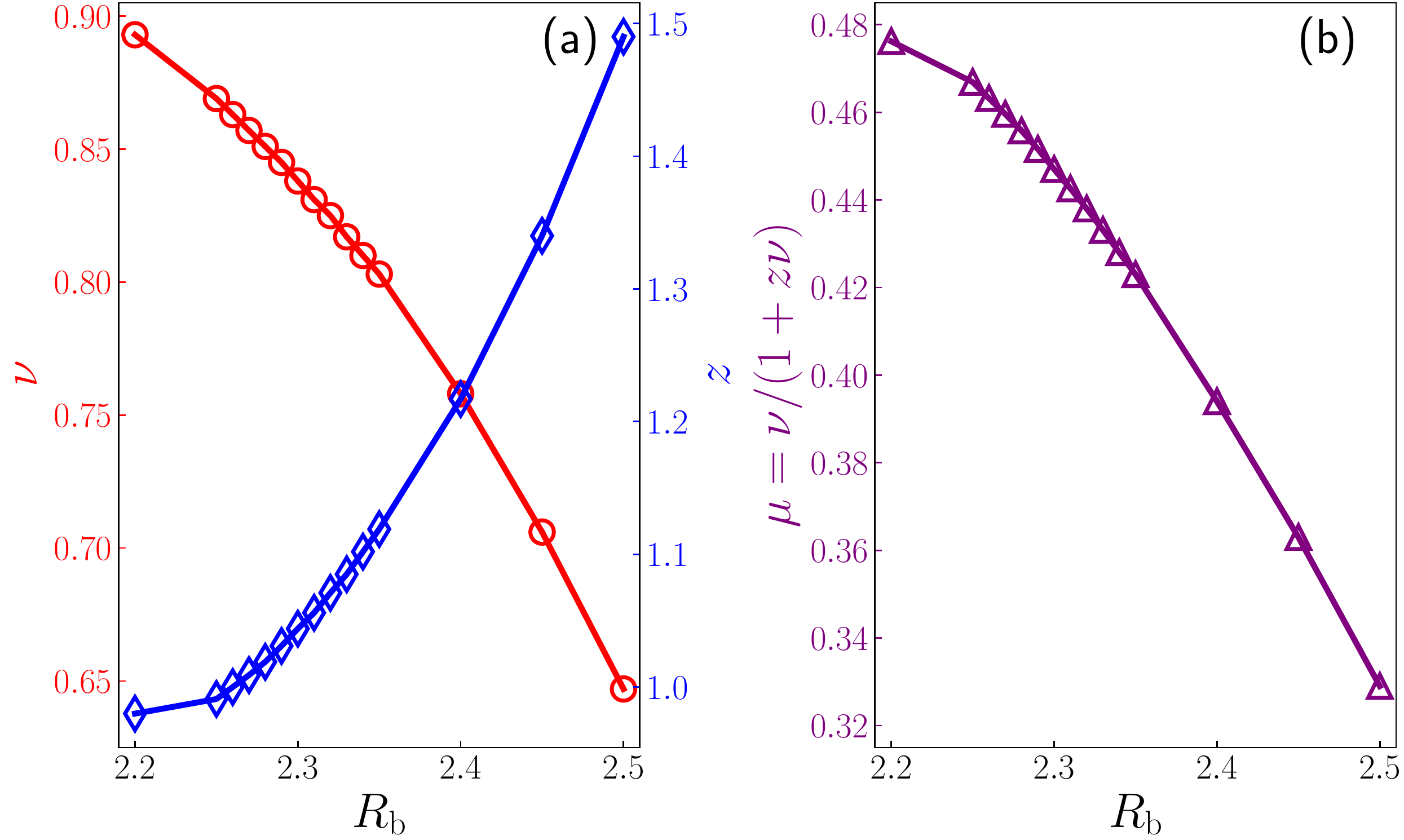}
\caption{
(a) The correlation length exponent $\nu$ (red circles) and dynamical critical exponent $z$ (blue diamonds) associated with the period-$3$ ordered-to-disordered transition for various Rydberg blockade radius (within the chiral transition region). The values of $\nu$ and $z$ are inferred from finite-size scaling analysis of $\chi_{F}$ and $\Delta$ with system sizes $L=82$ up to $127$\,. (b) The KZ exponent $\mu$ (purple triangles) obtained from the estimated $\nu$ and $z$ via the relation $\mu=\nu/(1+z\nu)$ is shown as a function of the blockade radius.
}
\label{fig:exponents}
\end{figure}

In Fig.~\ref{fig:exponents} and Table~\ref{tab:exponents}, we show the dependence of the exponents $\nu$, $z$, and $\mu$ as a function of the Rydberg blockade radius $R_{\rm{b}}$. The results show that these three critical exponents vary continuously with respect to the blockade radius in a monotonous manner, similar to the variation of the critical exponents of the chiral clock model reported in Ref.~\cite{samajdar2018pra}. More specifically, the correlation length exponent $\nu$ decreases with increasing blockade radius, while the dynamical critical exponent $z$ shows a steady increase with $\nu<\nu_{\rm{potts}}=5/6$ and $z>1$ for $R_{\rm{b}}/a\gtrsim2.3$\,. However, for the other side $R_{\rm{b}}/a<2.3$, we find that the exponent $z$ clearly has a value close to $1$. According to recent numerical works~\cite{maceira2022conformal,rader2019floating}, there should be a wide region belonging to the chiral universality below the Potts point. Therefore, we expect the critical exponent $z$ for $R_{\rm{b}}/a<2.3$ to be actually slightly larger than 1, which is hard to be identified in numerical calculations even when the potential finite-size effect is very small. To provide evidence that the transition indicated by blue diamonds in Fig.~\ref{fig:phase_diagram} (above or below the Potts point) is specifically in the chiral universality class, in Appendix~\ref{sec:chiralcheck}, we investigate the power-law behavior obeyed by the dominant wave vector approaching the commensurate value $2\pi/3$ as the ground state is driven from the disordered phase into the period-3 ordered phase for the case of $R_{\rm{b}}/a=2.4$ and $2.25$\,. The good data collapses displayed in Fig.~\ref{fig:chiralcheck} using the values of the critical exponents extracted from fidelity susceptibility and energy gap methods strongly support the existence of the chiral transition below the three-state Potts point. Therefore, our results are consistent with other numerical studies~\cite{maceira2022conformal,rader2019floating}. In addition, the value of the KZ exponent $\mu$ calculated here can also serve as a useful guide for further quantum KZ experiments.

\begin{ruledtabular}
\begin{table}[tb]
\caption{Critical exponents of the period-2 and -3 ordered-to-disordered transitions realized by Rydberg chains for different blockade radius $R_{\rm{b}}$ are summarized here for convenience. 
These exponents are extracted from finite-size scaling analyses of the fidelity susceptibility and energy gap.
}
\label{tab:exponents}
\begin{tabular}{lllll}
$R_{\rm{b}}/a$ 	& $\delta_{\rm{c}}$ 	& $\nu$ 	& $z$		& $\mu$ 		\\ \hline
1.4 			& 1.021(2) 		& 1.034 	& 0.9760		& 0.5146 		\\
1.6 			& 1.210(2) 		& 1.019 	& 0.9847		& 0.5086	 	\\
2.20 			& 1.962(3) 		& 0.893 	& 0.980			& 0.4762 		\\
2.25 			& 1.852(2) 		& 0.869 	& 0.991			& 0.4669 		\\
2.26			& 1.840(3)		& 0.863		& 1.000			& 0.4632		\\
2.27			& 1.829(3)		& 0.857		& 1.009 		& 0.4596		\\
2.28 			& 1.820(2) 		& 0.851 	& 1.019 		& 0.4558		\\
2.29 			& 1.813(3) 		& 0.845 	& 1.031 		& 0.4516		\\
2.30 			& 1.808(2) 		& 0.838		& 1.044			& 0.4470 		\\
2.31 			& 1.803(3) 		& 0.831 	& 1.056 		& 0.4426 		\\
2.32 			& 1.801(2) 		& 0.825 	& 1.071 		& 0.4380		\\
2.33 			& 1.799(3) 		& 0.817 	& 1.085			& 0.4331		\\
2.34 			& 1.799(2) 		& 0.810 	& 1.102 		& 0.4280		\\
2.35 			& 1.800(2) 		& 0.803 	& 1.119			& 0.4230 		\\
2.40 			& 1.818(2) 		& 0.758 	& 1.217			& 0.394	 		\\
2.45 			& 1.856(2) 		& 0.706 	& 1.34			& 0.363	 		\\
2.50 			& 1.914(2) 		& 0.647 	& 1.49			& 0.329
\end{tabular}
\end{table}
\end{ruledtabular}

\begin{figure}[tbp]
\includegraphics[width=\linewidth]{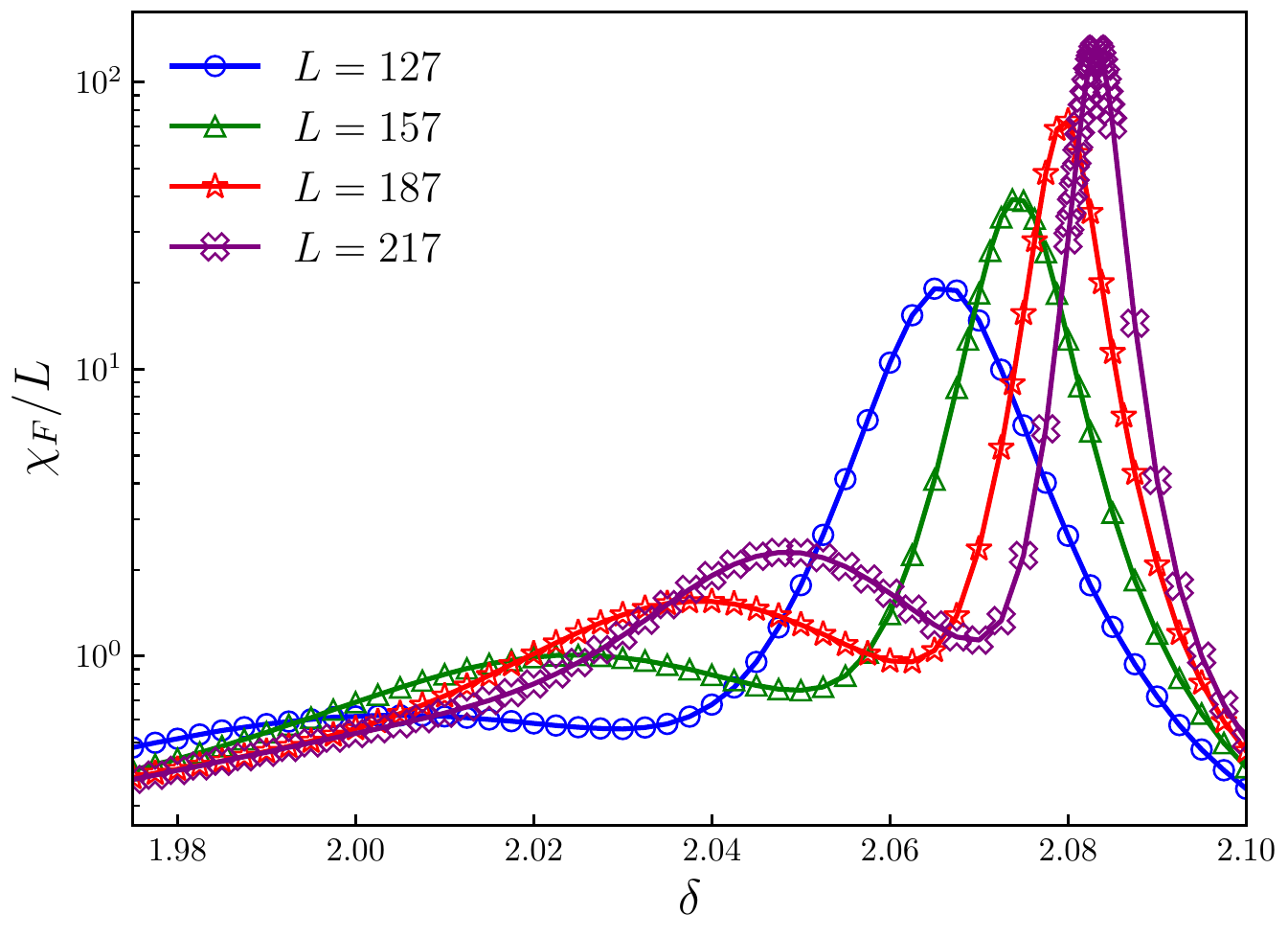}
\caption{
Fidelity susceptibility per site as a function of the detuning for the case of $R_{\rm{b}}/a=2.6$\,. $\chi_{F}/L$ shows two nearby peaks as the system is driven from the disordered phase into the period-3 ordered phase indicating the possible existence of an unconventional quantum phase between these two phases. The interval of these two peaks decreases as the system size is increased, which means the extent of the intermediate phase is very small, making its identification pretty hard.
}
\label{fig:double_peak}
\end{figure}

To complete the exploration of the period-3 ordered-to-disordered transition and to examine the recently reported floating phase, in Fig.~\ref{fig:double_peak}, we also investigate the case of $R_{\rm{b}}/a=2.6$, beyond the parameter range explored in Fig.~\ref{fig:exponents}.Surprisingly, the fidelity susceptibility per site now clearly shows two nearby peaks as a function of detuning $\delta$, implying that there may be another quantum phase different from the disordered or period-3 ordered one. Based on recent related works~\cite{chepiga2019prl,mila2021prr,maceira2022conformal}, we can expect this intermediate phase to be the so-called floating phase; as the system is driven from the disordered phase into the period-3 ordered phase, one experiences consecutively BKT and PT transitions. It is indeed that both of these susceptibility peaks become sharper with increasing chain length, and the floating phase is confined to a narrow parameter range, making its identification rather tough. 
In Appendix~\ref{sec:convergence}, we have also checked the effect of the MPS bond dimension on the fidelity susceptibility in the floating region and $\mathcal{D}=300$ is sufficient for the convergence.

\begin{figure}[tbp]
\includegraphics[width=1.\linewidth]{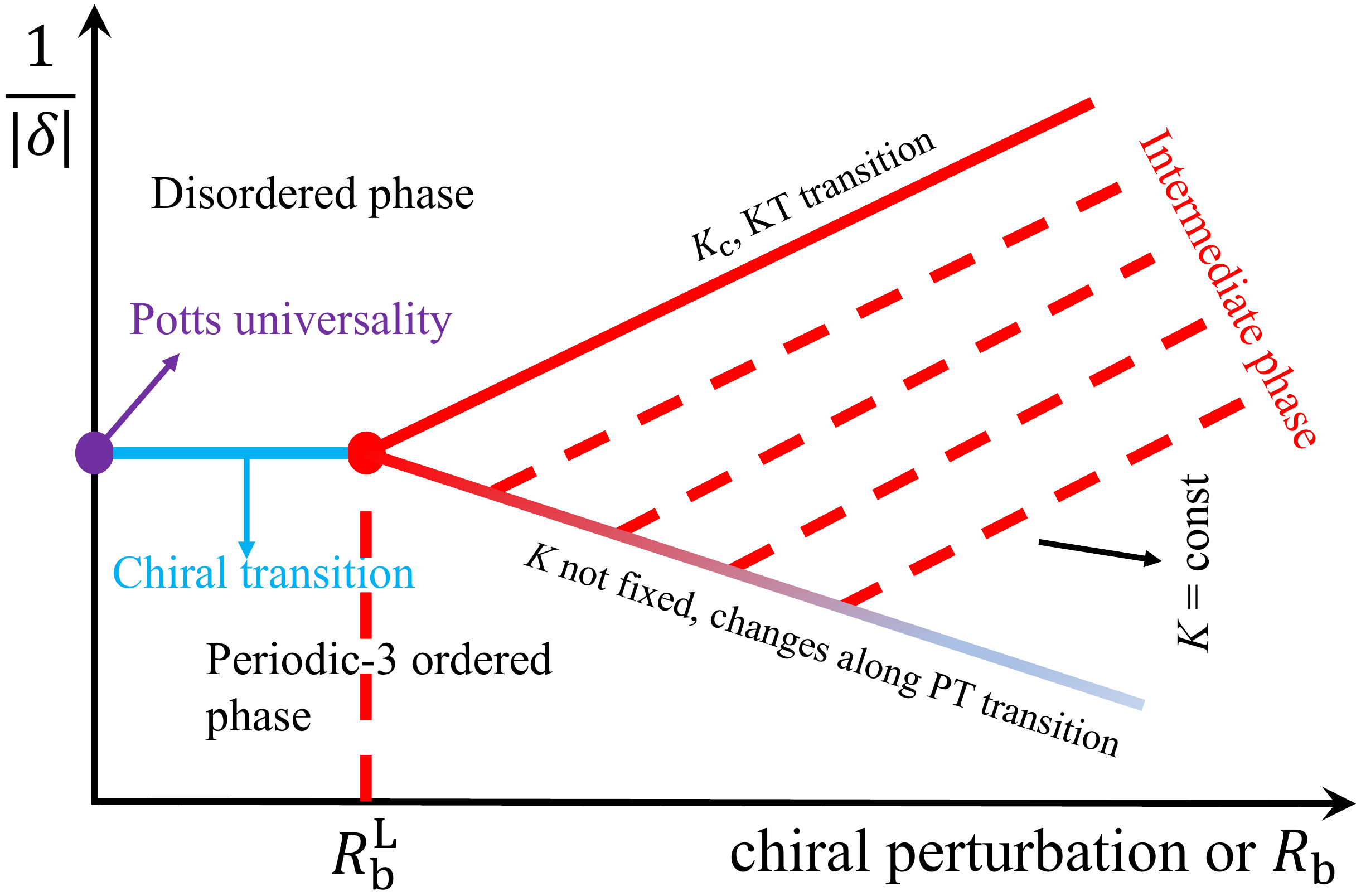}
\caption{The sketch of the physical picture of our numerical results for the Rydberg chain based on the argument proposed in Ref.~\cite{mila2021prr}. The Luttinger-liquid parameter $K$ changes along the PT transition line and reaches the value $K_{\rm{c}}$ of the KT transition at the Lifshitz point $R_{\rm{b}}^{\rm{L}}$ (the red point). Beyond this point, the phase transition is non-conformal chiral until it reaches the standard three-state Potts critical point (the purple point) where the chiral perturbation vanishes.
}
\label{fig:explanation}
\end{figure}

To summarize, with the numerical results and discussions shown above, we now have a clear picture of the period-3 C-IC transition consistent with the phase diagram mapped out in recent related works~\cite{rader2019floating,maceira2022conformal}. To further outline a general description of the physical picture, which is sketched in Fig.~\ref{fig:explanation}, we use the Luttinger-liquid concept and basically follow the argument proposed in Ref.~\cite{mila2021prr}. First, we note the striking fact that there are different types of domain walls concerning the commensurate period-3 ordering, and the generation of different domain-wall orderings may cost different energies~\cite{chepiga2021kibble}. This results in an asymmetry order of the domain walls, hence introducing a chiral perturbation in the low-energy field theory~\cite{whitsitt2018prb} (chiral means the asymmetry order of domain walls, e.g., $\rm{|A|B|C|}\ne\rm{|A|C|B|}$)
\begin{equation}
\begin{split}
\label{eq:action}
\mathcal{S}_{\Phi}=\int {\rm{d}}x {\rm{d}}\tau\, \big\{|\partial_{\tau}\Phi|^2+|\partial_x\Phi|^2+i\alpha_x\Phi^*\partial_{x}\Phi\,\,+\,\,\\
s_\Phi |\Phi|^2+u|\Phi|^4+\lambda\left[\Phi^p+(\Phi^*)^p\right]\big\}\,.
\end{split}
\end{equation}
Here $\Phi$ is the period-$p$ density-wave order parameter; $s_\Phi$, $u$, and $\lambda$ are the tuning parameter that drives the QPT, the interaction parameter, and perturbation that breaks the U(1) symmetry down to a $\mathbb{Z}_p$ one, respectively. The third term is induced by the chiral perturbation and drives the phase transition away from the standard Potts universality class to a chiral one. On the one hand, for the exact Potts point, the chiral perturbation happens to disappear. As the blockade radius $R_{\rm{b}}$ is increased or decreased from the Potts point, the energy-cost difference between the domain-wall orderings begins to appear (the chiral perturbation is induced), and the phase transition switches from the Potts universality into a chiral one. On the other hand, there is a gapless floating phase (characterized by the Luttinger-liquid parameter $K$) effectively having ``infinite'' inequivalent domain walls, located in the large $R_{\rm{b}}$ region (the chiral perturbation is strong). As the blockade radius is decreased, the domain walls will proliferate with an increasing $K$; the floating phase becomes unstable when the parameter $K$ reaches the BKT-transition value $K_{\rm{c}}$~\cite{mila2021prr}. Finally, the two sides merge at the Lifshitz point, whose existence is explained in Ref.~\cite{mila2021prr}, drawing a clear line between the chiral transition and the intermediate floating phase at the period-3 C-IC transition.

\section{Summary}
\label{sec:summary}
In conclusion, we perform large-scale finite-size DMRG simulations to investigate the ground-state phase diagram of the Rydberg chain in certain parameter regions. Using the concepts of fidelity susceptibility and energy gap as the diagnostic, we can efficiently locate quantum critical points between disordered and ordered phases of different density-wave orderings according to the blockade radius $R_{\rm{b}}$. For small values of $R_{\rm{b}}$, the phase transition between the period-2 ordered and disordered phases belongs to the (1+1)D Ising universality class, characterized by $\nu=1$ and $z=1$, which is consistent with previous numerical and experimental results. For intermediate values of $R_{\rm{b}}$, we found clear evidence of the continuous chiral transition between the period-3 ordered and disordered phases with non-conformal critical points. As a byproduct, the double-peak structure shown in the fidelity susceptibility also indicates the presence of an intermediate phase different from the conventional disordered or ordered phase, which is the gapless floating phase according to recent relevant works, for large values of $R_{\rm{b}}$. Our work demonstrates the potential advantage of using the fidelity susceptibility concept to detect this challenging critical phase. This paper shows that fidelity susceptibility can be used as an effective probe to study general C-IC transitions and provides a quantum information perspective for the understanding of non-conformal QPTs in programmable quantum simulators.

\begin{acknowledgments}
X.-J. Yu thank Zi-Xiang Li, Zhi Li, and Youjin Deng for helpful discussions. Numerical simulations were carried out with the ITensor package~\cite{itensor_paper}. We thank the computational resources provided by the TianHe-1A supercomputer, the High Performance Computing Platform of Peking University, as well as the Kirin No.2 High Performance Cluster supported by the Institute for Fusion Theory and Simulation (IFTS) at Zhejiang University. X.-J.Y. and L.X. is supported by the National Natural Science Foundation of China under Grant No. 11935002, and the National Key R\&D Program under Grant No. 2016YFA0300901. S.Y. and J.-B.X. is supported by the National Natural Science Foundation of China under Grant No. 11975198.
\end{acknowledgments}

\appendix

\section{More evidence for the (1+1)D Ising universality class}
\label{sec:isingmore}
In this appendix, we perform a finite-size scaling analysis of the fidelity susceptibility and energy gap for the case of $R_{\rm{b}}/a=1.4$ to give more evidence that the period-2 ordered-to-disordered transition hosted by the Hamiltonian $H_{\rm{Ryd}}$ belongs to the (1+1)D Ising universality class.

\begin{figure}[tbp]
\includegraphics[width=\linewidth]{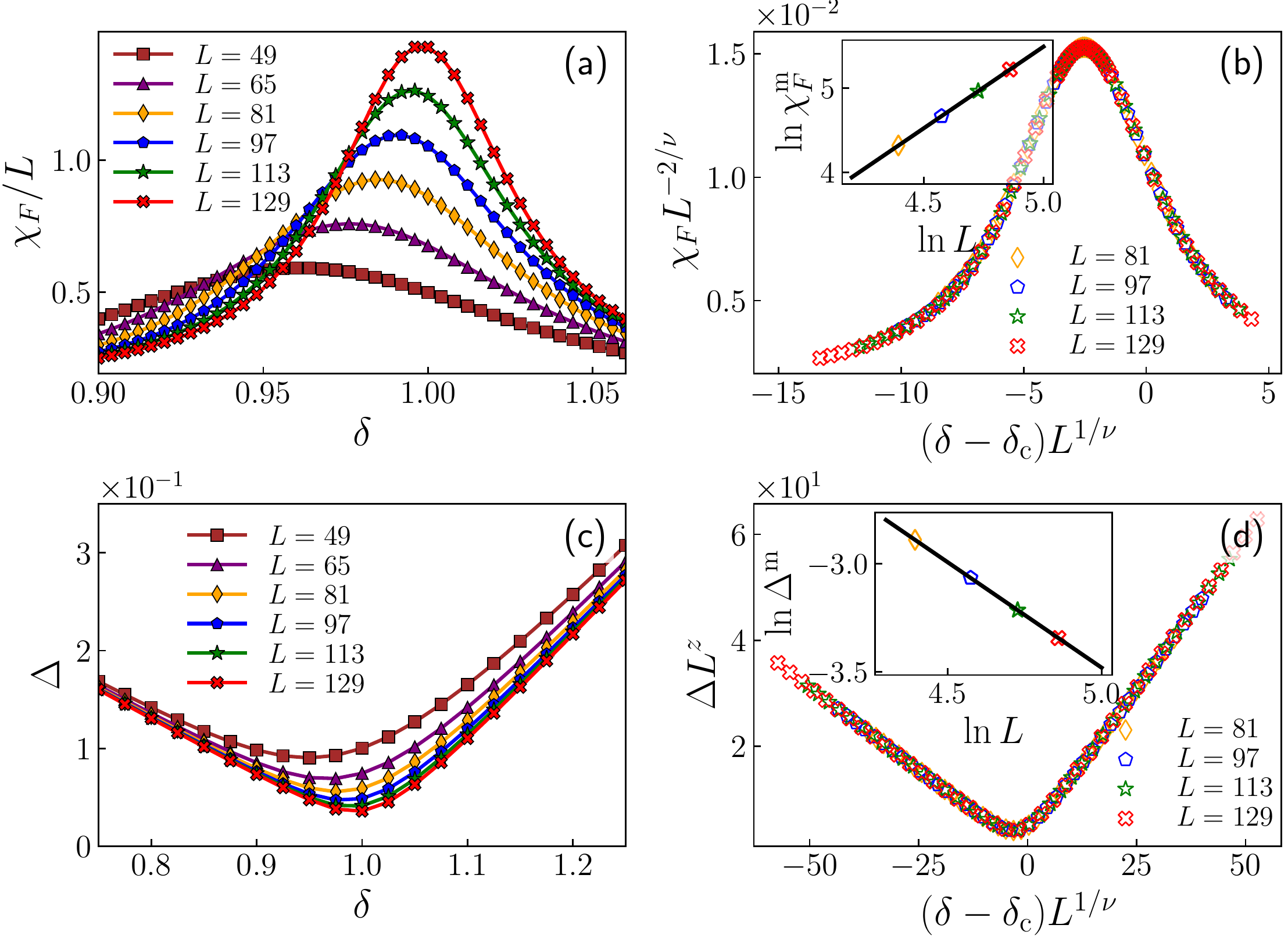}
\caption{
The finite-size scaling analysis of the fidelity susceptibility and energy gap for the period-2 ordered-to-disordered transition with a fixed blockade radius $R_{\rm{b}}/a=1.4$\,. The fidelity susceptibility per site $\chi_{F}/L$ (a) and energy gap $\Delta$ (c) develop a sharp peak and a deep valley, respectively, near the quantum critical point. A standard finite-size scaling analysis has been applied to estimate the critical point $\delta_{\rm{c}}=1.021(2)$\,, as well as the critical exponents $\nu=1.034$ and $z=0.9760$\,.
}
\label{fig:rb1.4}
\end{figure}

Using the same strategy adopted in Secs~\ref{sec:ising} and~\ref{sec:chiral}, in Figs.~\ref{fig:rb1.4}(a) and (c), we can clearly observe the fidelity susceptibility per site and energy gap develop, respectively, with an explicit peak and valley near a certain detuning indicating the transition from the disordered phase into the period-2 ordered one. According to finite-size scaling laws, the logarithms of $\chi_{F}$ and $\Delta$ are expected to exhibit linear dependence on $\ln{L}$; the corresponding least-square fittings are displayed in the insets of Figs.~\ref{fig:rb1.4}(b) and (d) from which we can extract the critical exponents $\nu=1.034$ and $z=0.9760$. With the obtained exponents, at last, we also achieve perfect curve collapses according to Eqs.~\eqref{eq:hamiltonian} and~\eqref{eq:ge_collapse} with a fine-tuned critical detuning $\delta_{\rm{c}}=1.021(2)$ as shown in Figs.~\ref{fig:rb1.4}(b) and (d).

Now we have investigated the phase transition at another Rydberg blockade radius $R_{\rm{b}}/a=1.4$. The estimation of the critical exponents extracted here also confirms the conclusion made in Sec~\ref{sec:ising} that the period-2 ordered-to-disordered transition in the programmable Rydberg chain belongs to the (1+1)D Ising universality class.

\section{Additional data collapses within the period-3 ordered-to-disordered transition regime}
\label{sec:z3more}
In this appendix, we show additional data collapses of the fidelity susceptibility and energy gap for the blockade radius within the period-3 ordered-to-disordered transition regime. 

\begin{figure*}[tbp]
\includegraphics[width=1.\linewidth]{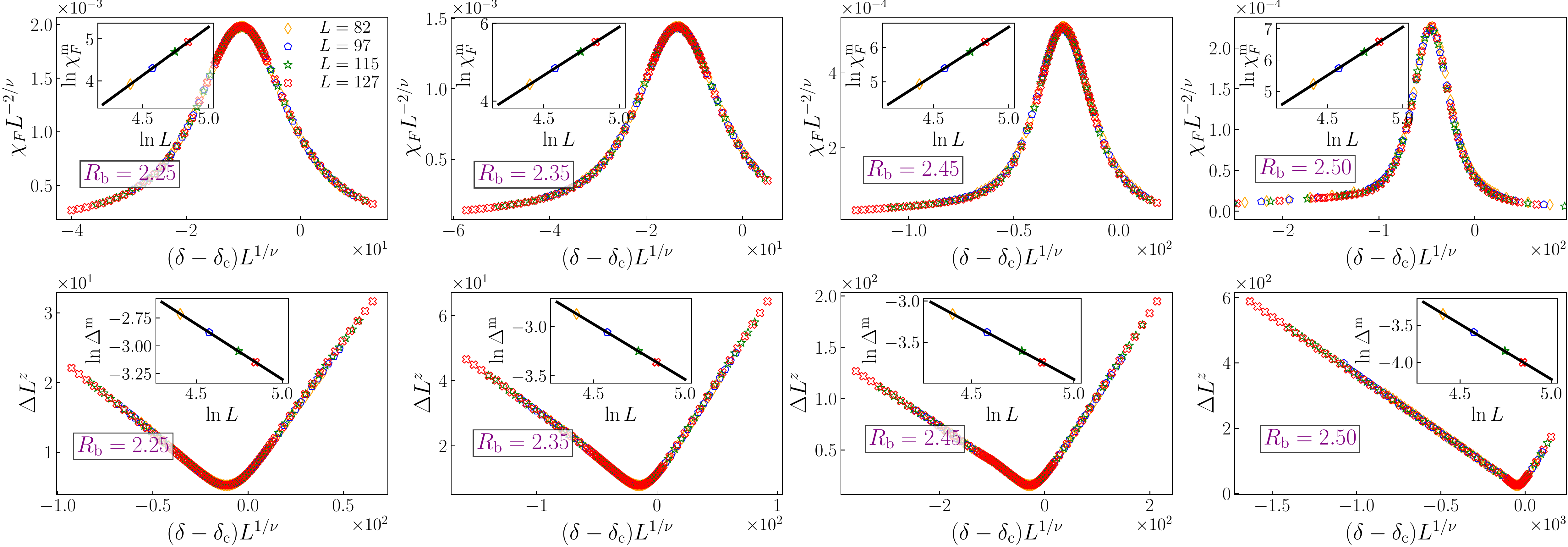}
\caption{
Data collapses of the fidelity susceptibility and energy gap for other blockade radius $R_{\rm{b}}$ within the period-3 ordered-to-disordered transition regime based on the values of $\nu$ and $z$ collected in Table~\ref{tab:exponents}. Different symbols are used to represent data for different system sizes: orange diamond for $L=82$, blue pentagon for $L=97$, green star for $L=115$, and red cross for $L=127$ (see the first plot). The critical exponents are extracted, following the same logic obeyed in Sec.~\ref{sec:ising} and~\ref{sec:chiral}, from the log-log plots of $\chi_{F}$ and $\Delta$ versus the chain length $L$ at the pseudocritical point (see the insets of the plots).
}
\label{fig:combine_two}
\end{figure*}

Following the same procedure illustrated in the main text, the numerical values of the critical exponents $\nu$ and $z$ are estimated by applying the finite-size scaling analysis. The fidelity susceptibility and energy gap for the case of $R_{\rm{b}}/a=2.25, 2.35, 2.45, 2.50$ are displayed in Fig.~\ref{fig:combine_two}, showing that the critical exponents vary continuously with respect to the blockade radius. It is noted that we have also performed the same analyses for the case of $R_{\rm{b}}/a=2.26, 2.27, \dots, 2.33, 2.34$; the data collapses are not shown here, but the extracted critical exponents are summarized in Table~\ref{tab:exponents}. 

\section{Evidence for the chiral transition}
\label{sec:chiralcheck}
To verify that the non-conformal phase transition detected in the main text is truly the chiral phase transition and belongs to the Huse-Fisher chiral universality class, we still have to resort to the diagnosis proposed originally by Huse and Fisher~\cite{huse1982prl}. Specifically, for the $p=3$ chiral phase transition, the product of the dominant wave vector $\lvert{q-2\pi/3}\rvert$ and the correlation length $\xi$ converges to a positive constant near the critical point, or equivalently, the exponent $\bar{\beta}$ describing the convergence of $q$ to $2\pi/3$ takes the same value as the exponent $\nu$.

\begin{figure}[tbp] 
\includegraphics[width=1.\linewidth]{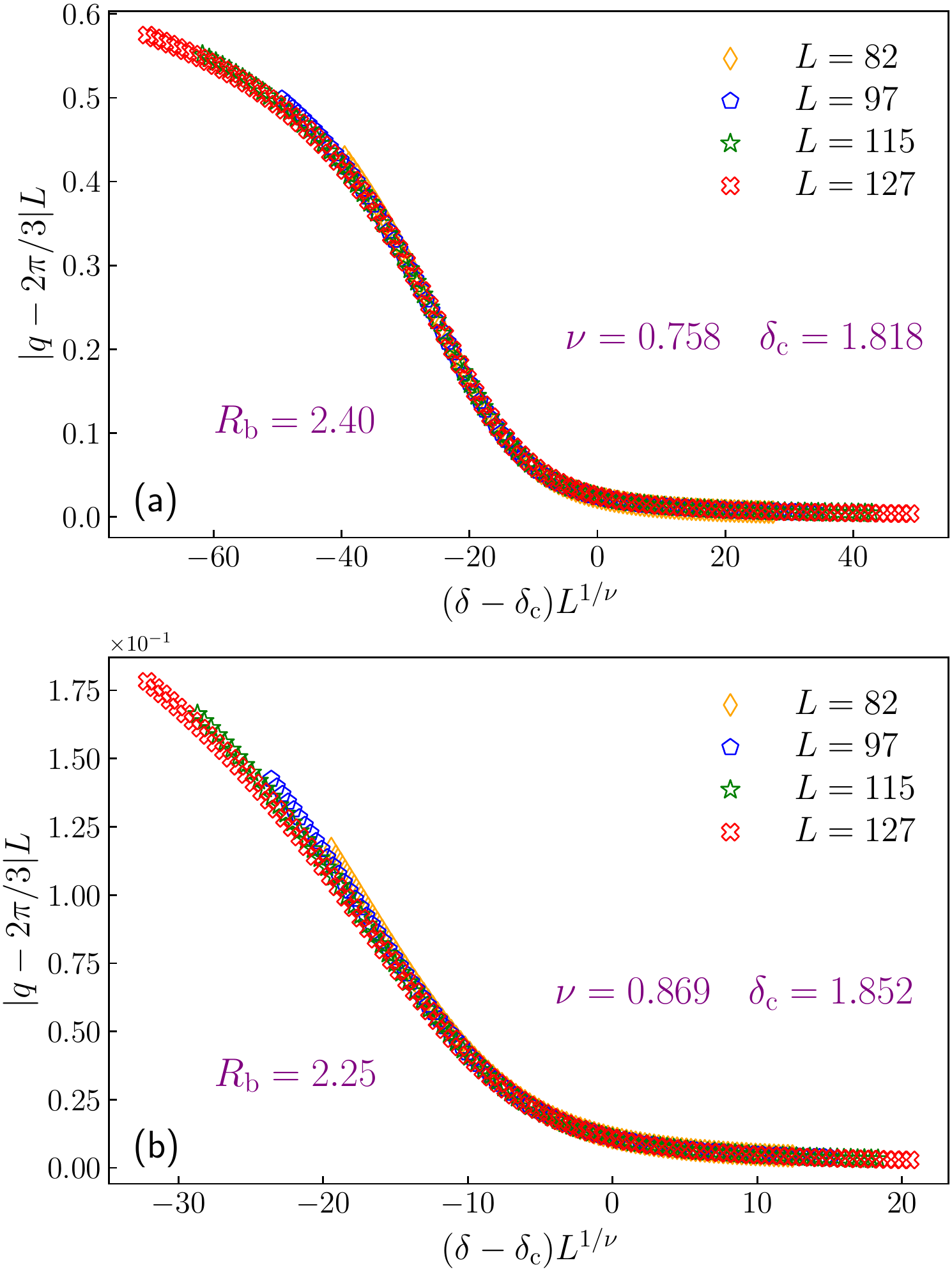} 
\caption{
Evidence for the chiral transition at the blockade radius $R_{\rm{b}}/a=2.4$ and $2.25$\,. The rescaled quantity $\lvert{q-2\pi/3}\rvert L$ is plotted as a function of $(\delta-\delta_{\rm c})L^{1/\nu}$ for different values of $L$ with $\delta_{\rm c}$ and $\nu$ extracted from the fidelity susceptibility approach (see Table~\ref{tab:exponents}). $\bar{\beta}=\nu$ has been assumed in this curve collapse [compared with Eq.~\eqref{eq:wave-vector}]. 
} 
\label{fig:chiralcheck} 
\end{figure}

As the ground state is driven into the ordered phase from the disordered phase, the dominant wave vector $q$ goes to $2\pi/3$ in a power-law behavior, $\lvert{q-2\pi/3}\rvert \sim (\delta-\delta_{\rm c})^{\bar{\beta}}$, characterized by the exponent $\bar{\beta}$. According to the finite-size scaling theory, we can then expect a universal functional form near the critical point, 
\begin{equation}
\lvert{q(\delta, L) - 2\pi/3}\rvert = L^{-\bar{\beta}/\nu} \mathcal{F}_{q}[L^{1/\nu}(\delta-\delta_{\rm c})]\,,
\label{eq:wave-vector}
\end{equation}
where $\mathcal{F}_{q}$ is an unknown scaling function. To obtain the value of the dominant wave vector $q(\delta, L)$ for finite systems, we calculate the density-density static structure factor $S(\delta, L)$ and $q(\delta, L)$ corresponds to the position of the maximum value of $S(\delta, L)$. In Fig.~\ref{fig:chiralcheck}, for the case of $R_{\rm{b}}/a=2.4$ and $2.25$, we plot $\lvert{q-2\pi/3}\rvert L$ as a function of $(\delta-\delta_{\rm c})L^{1/\nu}$ for different values of $\delta$ and $L$. By using the values of $\delta_{\rm c}$ and $\nu$ extracted from the fidelity susceptibility method in the plot, we can achieve good data collapses according to Eq.~\eqref{eq:wave-vector} without any free parameters. It is noted that $\bar{\beta}=\nu$ has been assumed in the curve collapses. The result confirms that the non-conformal phase transition observed in the main text is specifically in the Huse-Fisher universality class, and our finite-size scaling analyses are consistent with each other. We have also performed similar analyses for $R_{\rm{b}}/a=2.26, 2.27, 2.28, 2.29$ (the plots are not displayed here); our results support the chiral phase transition for these cases.

\section{Convergence of the fidelity susceptibility in the floating phase}
\label{sec:convergence}
\begin{figure}[tbp]
\includegraphics[width=.95\linewidth]{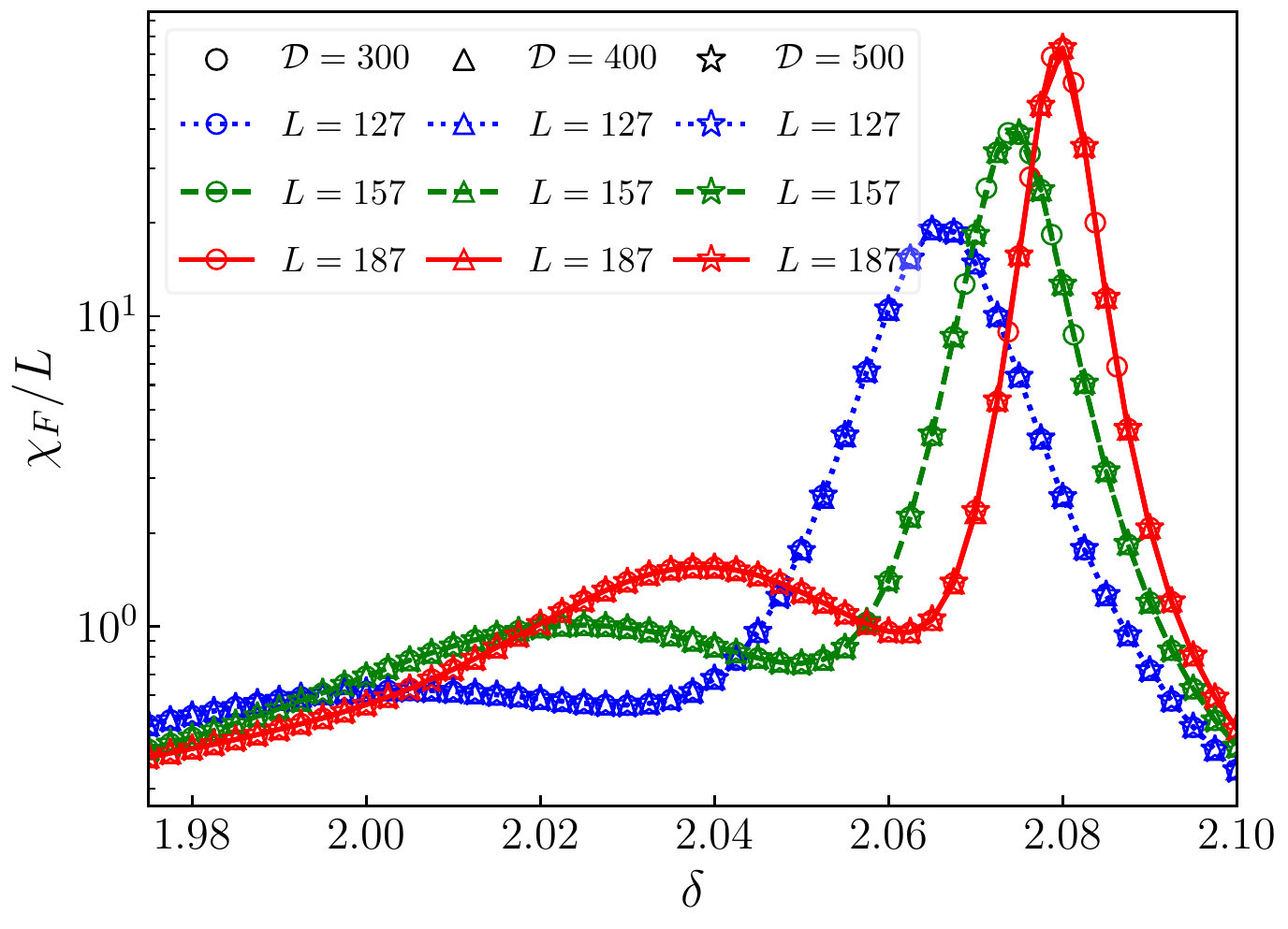}
\caption{
Convergence of the results of fidelity susceptibility in the floating phase for $R_{\rm{b}}/a=2.6$\,. $\chi_{F}/L$ is calculated for $L=127$ (dotted blue lines), $L=157$ (dashed green lines), and $L=187$ (solid red lines) with MPS bond dimensions $\mathcal{D}=300$ (circles), $\mathcal{D}=400$ (triangles), and $\mathcal{D}=500$ (stars).
}
\label{fig:convergence}
\end{figure}

In this appendix, we check the convergence of the fidelity susceptibility with respect to the MPS bond dimension in the floating phase. For this purpose, we calculate the values of $\chi_{\rm F}/L$ for $L=127$, $157$, and $187$ with several MPS bond dimensions $\mathcal{D}=300$, $400$, and $500$ at the blockade radius $R_{\rm b}=2.6$\,. As shown in Fig.~\ref{fig:convergence}, the MPS bond dimension $\mathcal{D}=300$ used in the main text has been sufficient to ensure the convergence. 

\section{Effect of the finite-system size}
\label{sec:finiteeffect}
\begin{figure}[tbp]
\includegraphics[width=1.\linewidth]{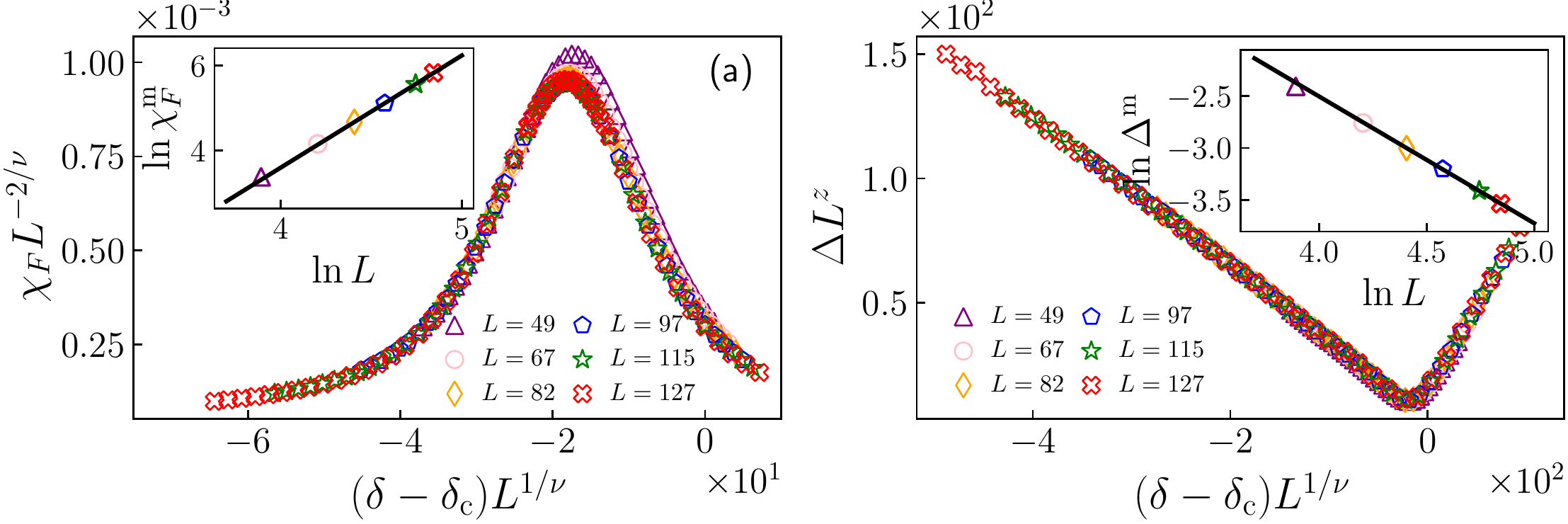}
\caption{
Data collapses of the rescaled $\chi_{F}$ and $\Delta$ for system size $L$ from $49$ to $127$ sites along the line $R_{\rm b}=2.4$\,. Fitting parameters: $\delta_{\rm c}=1.818$, $\nu=0.758$, and $z=1.217$ (see Fig.~\ref{fig:combine_one}). The insets are the linear log-log plots of $\chi_{F}^{\rm m}$ and $\Delta^{\rm m}$ with respect to $L$ (only the data points of largest four sizes are used in the fitting process).
}
\label{fig:datacollapse}
\end{figure}

In this appendix, we investigate the effect of the finite system sizes on the extraction of the critical properties, namely, the associated exponents $\nu$ and $z$. To this end, in Fig.~\ref{fig:datacollapse}, we perform data collapses of rescaled $\chi_{F}$ and $\Delta$ for system size $L$ from $49$ to $127$ sites along the horizontal line $R_{\rm{b}}/a=2.4$ through the chiral transition.

The values of the critical exponents $\nu=0.758$ and $z=1.217$, as well as the critical point location $\delta_{\rm c}=1.818$, used in the curve collapses are the results from Fig.~\ref{fig:combine_one}. It is clear that, for both $\chi_{F}$ and $\Delta$, the data collapse of the largest four system sizes are quite good with only slight deviation of the smallest two sizes. In addition, the linear log-log fitting of the maximum (minimum) value of the fidelity susceptibility $\chi_{F}^{\rm m}$ (energy gap $\Delta^{\rm m}$) versus the system size $L$ shown in the insets of Fig.~\ref{fig:datacollapse} also indicates the fact that the system sizes $L$ from $82$ to $127$ sites are sufficient to estimate accurate critical exponents for the chiral transition. Therefore, we can expect that the information summarized in Table~\ref{tab:exponents} are reliable to represent the results in the thermodynamic limit.

\bibliography{rydberg_chain_arxiv}
\end{document}